\documentclass[prl,amsmath,superscriptaddress,twocolumn,showpacs]{revtex4-1}
\usepackage{bm}
\usepackage{amssymb}
\usepackage{colordvi}
\usepackage{graphicx}
\usepackage{color}
\usepackage{hyperref}
\newcommand{\up}{\uparrow}
\newcommand{\down}{\downarrow}
\def\spin{{\mbox{\boldmath{$\hat{\sigma}$}}}}
\newcommand{\ov}[1]{\overline{{#1}}}
\newcommand{\be}{\begin{equation}}
\newcommand{\ee}{\end{equation}}
\newcommand{\bea}{\begin{eqnarray}}
\newcommand{\eea}{\end{eqnarray}}

\newcommand{\vep}{\varepsilon}
\newcommand{\ave}[1]{\langle #1\rangle}

\def\nn{\nonumber}

\begin{document}



\title{Emergence of triplet orbital pairing and non-Abelian states in
  ultracold multi-orbital optical lattices with quadratic band touching} 

\author{Jian-Hua Jiang}
\email{jianhua.jiang@weizmann.ac.il}
\affiliation{Department of Condensed Matter Physics, Weizmann Institute of
  Science, Rehovot 76100, Israel}

\date{\today}

\begin{abstract}
It is found that all the {\em singlet orbital pairing} instabilities
are {\em absent} in a class of spin-polarized
multi-orbital systems with quadratic band touching, which opens the
way for {\em triplet orbital pairing} order. The ground states are
found to be {\em non-Abelian} states with $p$-wave orbital pairing in
checkerboard (away from 1/2 filling) and kagome (above 1/3 filling)
lattices with {\em isotropic} attractive interaction which can be realized
in ultracold multi-orbital optical lattices. The special
property of such systems is generalized to more classes of
multi-orbital systems, where the fully-gapped {\em non-Abelian} states
are possibly the ground states. Those findings are helpful in
achieving topological quantum computation.
\end{abstract}







\maketitle

{\sl Introduction.}-- Excitations obeying non-Abelian statistics can
emerge in interacting many-fermion systems\cite{MR,Kitaev}. One known
prototype is the fully-gapped $p$-wave superconducting (SC)/superfluid
(SF) state \cite{ReadGreen}. More generally, any fully-gapped SC/SF state
with {\em odd} Chern number is a {\em non-Abelian} state, as there is one
topologically protected zero-energy Majorana bound state in each 
quantized vortex and the braiding of the Majorana fermions leads to
the non-Abelian statistics\cite{Ivanov}. The search for systems with
non-Abelian excitations is desirable not only in the sense that they
can be used to realize topological quantum
computation (TQC) \cite{Kitaev,tqc,Nayak,atom-qc}, but also that they have
nontrivial ground states which are characterized by topological
orders\cite{Wen}.

Except in a few cases\cite{TI,soc,QiQAH,probing,our}, triplet pairing
is crucial to the emergence of non-Abelian states. However, in reality
triplet pairing is scarce, whereas singlet pairing prevails. One of the
reason is that the interaction between fermions with opposite spin
(e.g., on-site interaction) is stronger than that between electrons
with the same spin (e.g., nearest neighbor interaction) due to Pauli
exclusion. In spinless (spin-polarized) fermionic systems, the
situation is different. In lattice systems with a single orbit (site)
in an unit cell, only the triplet pairing is possible. However, in
multi-orbital systems, where pseudo-spin denotes the orbital degree of
freedom, the singlet (inter-orbital) pairing prevails due to similar
reasons.

In this Letter, we propose a scenario that suppresses {\em all} the
{\em singlet} (inter-orbital) pairing instabilities and opens the way to the
{\em triplet} orbital pairing, regardless of the relative strength of the
pairing interaction of the two. The concerned systems have a single
quadratic band touching (QBT) protected by time-reversal and
space-inversion symmetry\cite{sun2}, while the Fermi level is above or
below the QBT point. The unique property of such systems is that the
pseudo-spin polarization on the Fermi surface has a winding number of
$\pm 2$. Due to such winding as well as the time-reversal symmetry,
the ${\bf k}$ and $-{\bf k}$ states on the Fermi surface has the same
pseudo-spin polarization (pseudo-spin polarization winds one period
when ${\bf k}$ winds to $-{\bf k}$), while the states with opposite
pseudo-spin is well below the Fermi surface. Hence there is {\em no}
singlet orbital pairing instability in the weak pairing regime, which
opens a way to {\em triplet orbital pairing orders} in multi-orbital
systems. For concreteness, we study two systems with a single QBT: the
checkerboard (away from 1/2 filling) and kagome (above 1/3 filling)
lattices with {\em isotropic} attractive interaction. It is found
that the ground states in those systems are {\em non-Abelian} states
with $p$-wave orbital pairing which is promising for TQC. Furthermore,
the special property of such systems is generalized to more classes of
multi-orbital systems, where the {\em non-Abelian} states which have
fully-gapped Fermi surface to gain more condensation energy
are possibly the ground states\cite{Anderson,Cheng}.

Ultracold fermionic atom/polar-molecule gases with tunable
interaction through Feshbach resonance\cite{fesh,bfmix}
and other techniques\cite{m} offers a lot of advantages in
realizing generic interacting many-fermion
systems\cite{rmp}. Especially, in polar-molecule systems
the combination of microwave excitation with
dipole--dipole interactions enables a variety of effective inter-molecule
interactions in a designable fashion and with significant strength to
achieve observable emergent phases\cite{m}. The good
controllibility and emerging new detection techniques also
enable them to be an ideal platform to achieve TQC\cite{atom-qc},
given that the non-Abelian states can be realized. We show that the
systems with non-Abelian ground states proposed in this Letter can be
realized in spin-polarized ultracold fermionic systems in
multi-orbital optical lattices\cite{multi,ch,ka,sun2}. With the emerging
technology advancements in multi-orbital optical lattices\cite{multi}
and ultracold fermionic atom/polar-molecule gases\cite{m}, the
proposed systems are helpful in achieving TQC.

\vskip 0.15cm
{\sl Quadratic band touching in checkerboard and kagome lattices.}--
The checkerboard lattice is depicted in Fig.~1(a). In each unit cell
there are two sites labeled as red (circle) and blue (square) dots in
the figure. Allowing one orbit in each site, for spin-polarized
fermions, the Hamiltonian takes the form
\be
H = - \sum_{\langle i\sigma,j\sigma^\prime\rangle}
t_{i\sigma,j\sigma^\prime}(c^\dagger_{i\sigma}c_{j\sigma^\prime}+H.c.)
+ H_{\rm int} .
\ee
Here $i$ and $j$ are the indices of the unit cells, while pseudo-spins
($\sigma,\sigma^\prime=\up,\down$) denote the two different orbits
in each unit cell. $\langle i\sigma,j\sigma^\prime\rangle$ restricts
the summation to the nearest and next nearest neighbors. The system is
engineered in such a way that the hopping amplitude in $x$ and $y$
directions between red (blue) sites are $t^\prime$ [solid links in
  Fig.~1(a)] and $t^{\prime\prime}$ [dotted links] ($t^{\prime\prime}$
and $t^{\prime}$) respectively. The hopping between red and blue sites
are $t$ [dashed links]. Accordingly, the free Hamiltonian is
$H_0 = \sum_{\ov{{\bf k}}}\psi_{\ov{{\bf k}}}^\dagger{\cal
  H}_0(\ov{{\bf k}})\psi_{\ov{{\bf k}}}$ with $\psi_{\ov{{\bf
      k}}}=(c_{\ov{{\bf k}}\up},c_{\ov{{\bf k}}\down})^T$, where
\be
{\cal H}_0(\ov{{\bf k}}) = h_0(\ov{{\bf k}})\sigma_0 + {\bf h}(\ov{{\bf k}})
\cdot\spin
\label{h0k}
\ee
where $\hat{\sigma}_0$ denotes the $2\times 2$ identity matrix
and $\spin$ is the Pauli matrix vector. $h_0(\ov{{\bf k}})=-2t_0(\cos \ov{k}_x +
\cos \ov{k}_y)$, $h_z(\ov{{\bf k}})=-2t_z(\cos \ov{k}_x - \cos
\ov{k}_y)$, and 
$h_x(\ov{{\bf k}})= 8t_x\cos\frac{\ov{k}_x}{2}\cos\frac{\ov{k}_y}{2}$ with 
$t_0=(t^\prime+t^{\prime\prime})/2$,
$t_z=(t^\prime-t^{\prime\prime})/2$, $t_x=-t/2$. $h_y(\ov{{\bf k}})=
0$ due to the time-reversal and space-inversion
symmetry\cite{Sun}. Due to the symmetry, the single QBT point can only
be at a time-reversal invariant momentum ${\bf K}=-{\bf K}$. The QBT
is a ${\bf k}$-space vortex which is topologically
stable\cite{sun2}. In checkerboard lattice the two bands 
touch quadratically [Fig.~1(c)] at ${\bf K}=(\pi, \pi)$\cite{Sun}.
In the vicinity of ${\bf K}$, one has $h_0({\bf k})=t_0k^2$, $h_z({\bf
  k})=t_z(k_x^2-k_y^2)$, $h_x({\bf k})=2t_xk_xk_y$, and $h_y({\bf k})
= 0$, where ${\bf k}=\ov{{\bf k}}-{\bf K}$. As ${\bf K}$ is a time-reversal
invariant momentum, pairing is between the ${\bf k}$ and $-{\bf k}$
states. The spectrum is $\vep_{{\bf k}\pm}=t_0k^2\pm
k^2\sqrt{t_z^2\cos^2(2\theta_{\bf k})+t_x^2\sin^2(2\theta_{\bf k})}$
with $\theta_{\bf k}={\rm Arg}[k_x+ik_y]$. At half-filling, the Fermi
level is at the QBT point. Away from it only one band crosses the
Fermi level [Fig.~1(c)] when $|t_0|\le |t_z|,|t_x|$. A characteristic
of such systems is that the pseudo-spin polarization on the Fermi
surface has a winding number of $\pm 2$. The winding number is
\be
N_w = \frac{1}{2\pi}\oint_{\rm FS} d\phi_{\bf k} ,
\ee
where ${\rm FS}$ stands for the Fermi surface and $\phi_{\bf k} = {\rm
  Arg}[h_z({\bf k})+ih_x({\bf k})]$ is the direction of the
pseudo-spin polarization in the $z$-$x$ plane. Winding number
$N_w=2{\rm sgn}(t_xt_z)=\pm 2$ as well as the time-reversal symmetry
guarantee that the pseudo-spin polarization at ${\bf k}$ is the same
as that at $-{\bf k}$ on the Fermi surface [Fig.~1(d)].

In the kagome lattice, there are three different sites in each unit
cell, labeled as red (circle), blue (square) and green
(triangle) in Fig.~1(b), which we denote as 1, 2, and 3. With only the
nearest neighbor hopping, the free Hamiltonian can be written as $H_0
= \sum_{{\bf k}}\psi_{\bf k}^\dagger{\cal H}_{\rm kgm}({\bf k})\psi_{\bf k}$
where $\psi_{\bf k}=(c_{{\bf k}1},c_{{\bf k}2},c_{{\bf k}3})^T$ and
\be
{\cal H}_{\rm kgm}({\bf k}) = -2t \left( \begin{array}{cccc}
  0 & \cos \frac{k_{12}}{2} & \cos \frac{k_{13}}{2} \\
  \cos \frac{k_{12}}{2} & 0 & \cos \frac{k_{23}}{2} \\
  \cos \frac{k_{13}}{2} & \cos \frac{k_{23}}{2} & 0 \\
\end{array}\right)  \nn
\ee
with $t$ being the hopping amplitude. $k_{ij}={\bf k}\cdot
{\bf n}_{ij}$ for $i,j=1,2,3$ with ${\bf n}_{12}=(1,0)$, ${\bf
  n}_{13}=(\frac{1}{2},\frac{\sqrt{3}}{2})$ and 
${\bf n}_{23}={\bf n}_{13}-{\bf n}_{12}$. The spectrum is
$E_{{\bf k}0}=2t$, $E_{{\bf k}\pm}=-t\pm
t\sqrt{4(\cos^2\frac{k_{12}}{2}+\cos^2\frac{k_{13}}{2}+\cos^2\frac{k_{23}}{2})-3}$.
In kagome lattice the $+$ and $0$ bands touch quadratically at ${\bf
  K}=(0,0)$. Around ${\bf K}$, the spectrum is approximately $E_{{\bf
    k}+}=2t-\frac{1}{4}tk^2$, $E_{{\bf k}0}=2t$, and $E_{{\bf
    k}-}=-4t+\frac{1}{4}tk^2$. For $t<0$ and not much above $1/3$
filling (the results are similar for $t>0$ below $2/3$ filling), only
the $+$ band crosses the Fermi surface. Projecting out the $-$ band
which is far above, one obtains an effective Hamiltonian
in the form of Eq.~(\ref{h0k}) with $h_{0}({\bf k}) =
\frac{1}{8}|t|k^2$, $h_z({\bf k}) =\frac{1}{8}|t|(k_x^2-k_y^2)$,
$h_x({\bf k}) =\frac{1}{4}|t|k_xk_y$, and $h_y({\bf k}) = 0$, where
the pseudo-spin up and down states are defined as
$|\up\rangle=\frac{1}{\sqrt{2}}(1,-1,0)^T$ and
$|\down\rangle=\frac{1}{\sqrt{6}}(1,1,-2)^T$ respectively.

\begin{figure}[htb]
\includegraphics[height=2.8cm]{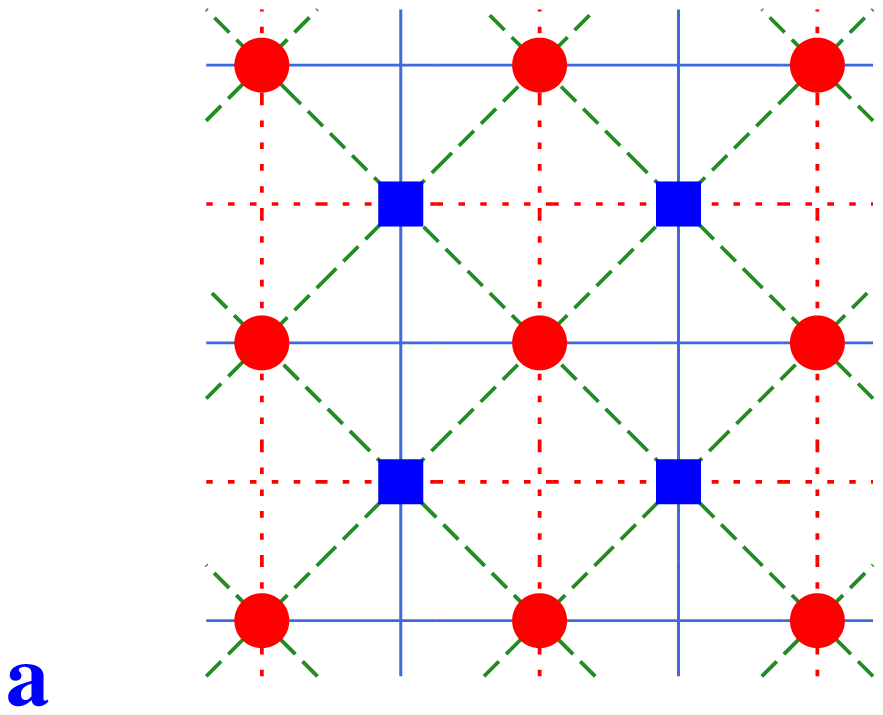}\includegraphics[height=2.8cm]{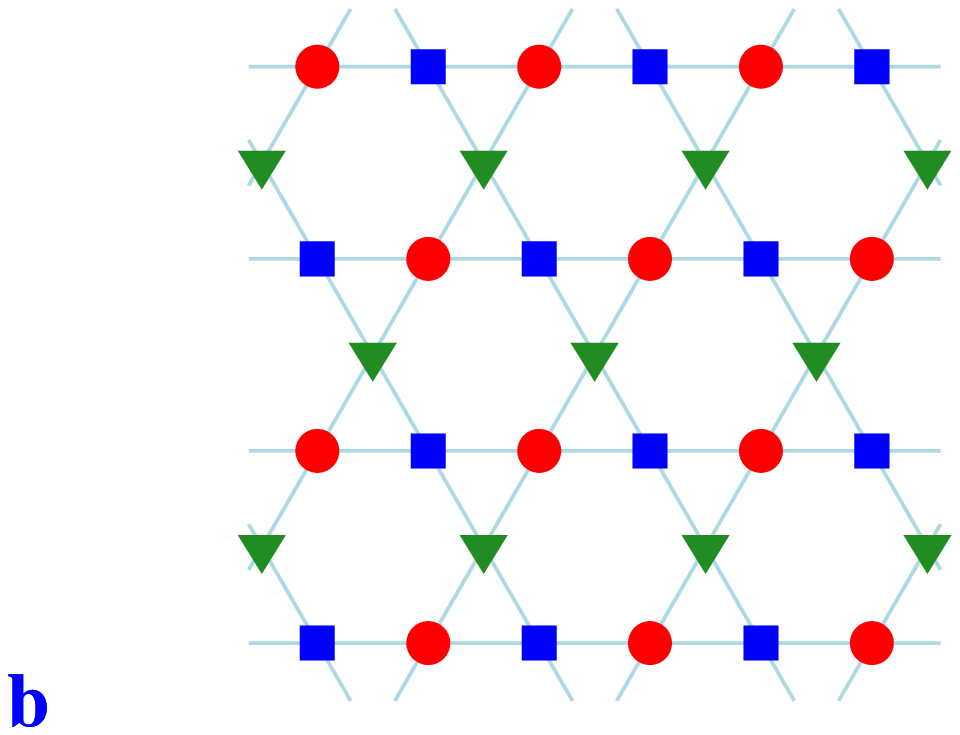}
\includegraphics[height=2.8cm]{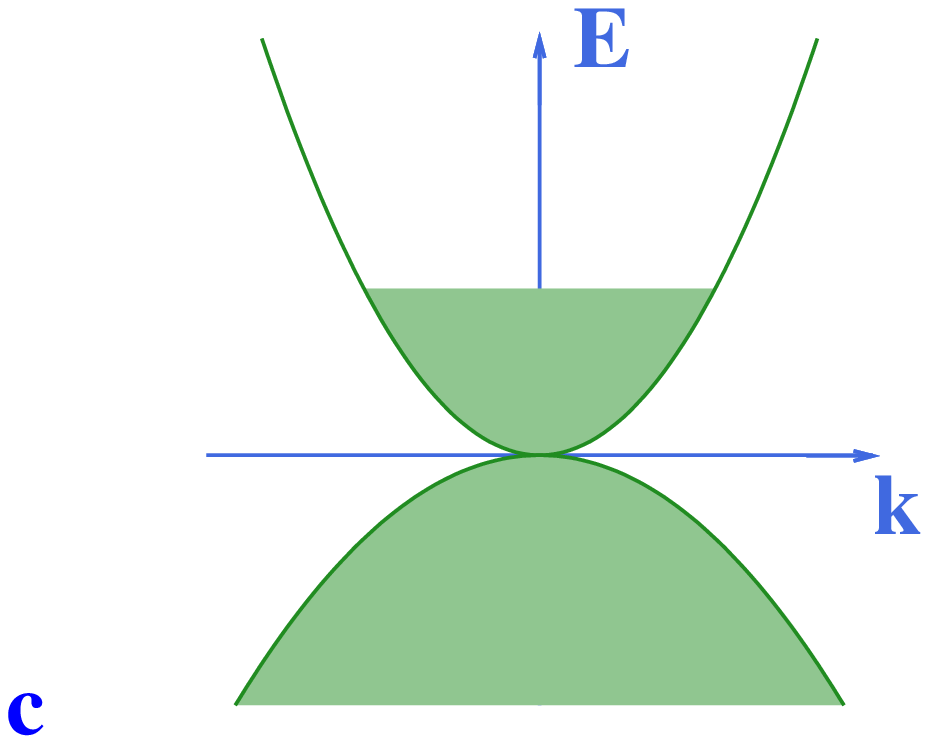}\includegraphics[height=2.8cm]{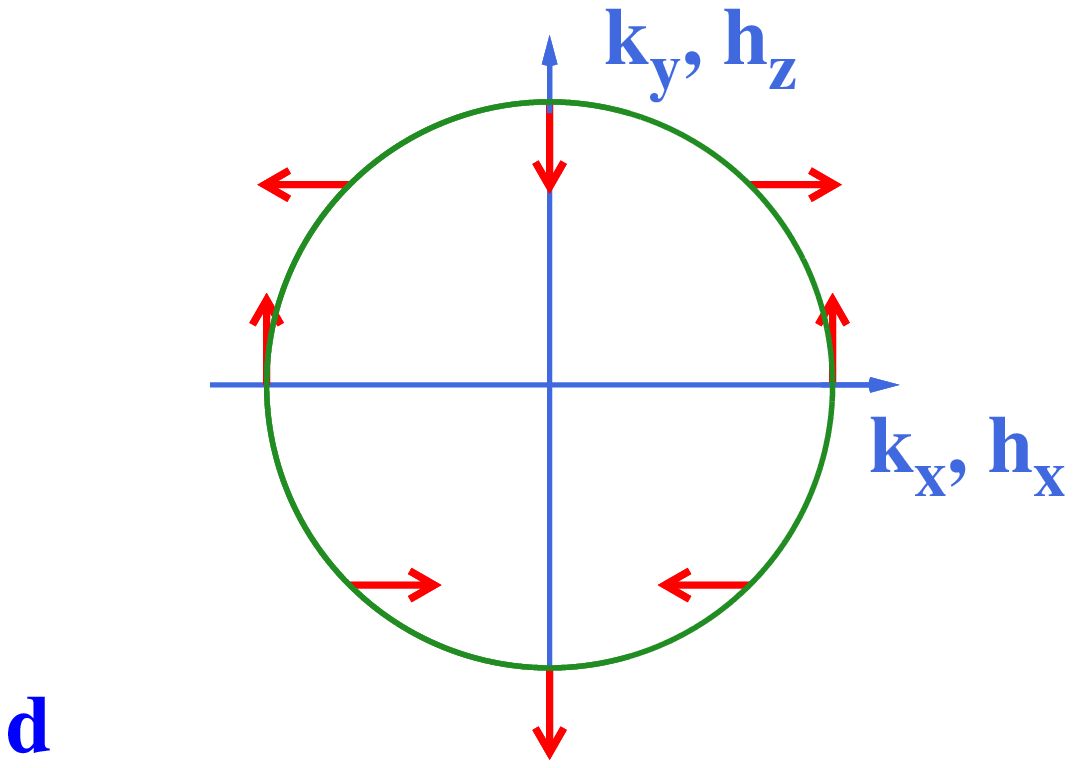}
\caption{(Color online) (a) Checkerboard and (b) kagome lattices.
  In checkerboard lattice, the hopping amplitude along the blue
  (solid), red (dotted), and green (dashed) links are $t^\prime$,
  $t^{\prime\prime}$, and $t$ respectively. (c) Band structure and
  band filling near half-filling with $t_x=t_z$ in the checkerboard
  lattice. ${\bf k}$ is measured from the QBT point ${\bf K}$. (d)
  Direction of the pseudo-spin field $(h_x,h_z)$ (also represents the
  pseudo-spin polarization direction) on the Fermi surface of a QBT
  system.}
\label{fig1}
\end{figure}

\vskip 0.15cm
{\sl Nonexistence of singlet  pairing instability in QBT systems.}-- 
Diagonalizing the Bogoliubov-de Gennes (BdG) Hamiltonian with singlet
pairing, $H_{\rm int}^{\rm sp}=-\frac{1}{2}\sum_{\bf
  k}\Delta_s({\bf k})(c_{{\bf k}\up}^\dagger c_{-{\bf k}\down}^\dagger - c_{{\bf
    k}\down}^\dagger c_{-{\bf k}\up}^\dagger)+{H.c.}$, in QBT systems,
one finds that the spectrum is {\em gapless}. The gapless feature is
most transparent when $h_{0}({\bf k})= 0$, where the spectrum is
$\pm \left[\vep_{{\bf k}\pm} -\sqrt{\mu^2+|\Delta_s({\bf
      k})|^2}\right]$ with $\mu$ being the chemical potential. For
such spectrum, there is {\em no} pairing instability in the weak
pairing regime.

\vskip 0.15cm
{\sl Interaction and $p$-wave pairing.}-- Consider the triplet
pairing instabilities due to {\em isotropic} attractions between
fermions at nearest and next nearest neighbor sites,
\be
H_{\rm int} = - \frac{1}{2}\sum_{<i\sigma,j\sigma^\prime>}
V_{\sigma\sigma^\prime}n_{i\sigma} n_{j\sigma^\prime} .
\ee
The physical realization of such interaction will be discussed in the
end of the paper. In checkerboard lattices, we denote $V_{\up\down}=V$
and $V_{\up\up}=V_{\down\down}=U$.
Using a Hubbard-Stratonovich decoupling with BCS pairing and ignoring
superconducting fluctuations, one gets
\bea 
H_{\rm int}^{\rm MF} &=&  - \frac{1}{2} \sum_{{\bf k}{\nu}}
\Bigl[\Delta_\nu({\bf k}) d_\nu({\bf k}) + H.c. \Bigr] + \frac{1}{2}E_0,
\eea
where $d_{x/y}({\bf k})=c_{{\bf k}\up}^\dagger c_{-{\bf
      k}\up}^\dagger \mp c_{{\bf k}\down}^\dagger c_{-{\bf
      k}\down}^\dagger$, $d_z({\bf k})=c_{{\bf k}\up}^\dagger c_{-{\bf
      k}\down}^\dagger+c_{{\bf k}\down}^\dagger c_{-{\bf
      k}\up}^\dagger$, and $E_0=\sum_{{\bf  k}\nu}\Delta_\nu({\bf
  k})\ave{d_\nu({\bf k})}$. The triplet  pairing is
$\Delta_\nu({\bf
  k}) = V_\nu\sum_{\bf p}S({\bf k},{\bf p})\ave{d_\nu^\dagger({\bf
    p})} $,
where $V_x=V_y=U$, $V_z=V/2$, and $S({\bf k},{\bf p}) \simeq
\frac{1}{2}(k_+p_-+k_-p_+)$ with $k_{\pm}=k_x\pm ik_y$. Hence the
pairing is $p$-wave type.

For kagome lattice with {\em isotropic} nearest neighbor attractive
interaction, one obtains a similar Hamiltonian with
$\Delta_y=-\frac{1}{3}(\Delta_{12}+\Delta_{13}+\Delta_{23})$,
$\Delta_x=-\frac{1}{3}(2\Delta_{12}-\Delta_{13}-\Delta_{23})$,
$\Delta_z=\frac{1}{\sqrt{3}}(\Delta_{23}-\Delta_{13})$. Here
$\Delta_{ij}=\frac{V^\prime}{4}\sum_{\bf p}\ave{c_{-{\bf p}i}c_{{\bf
      p}j}+c_{-{\bf p}j}c_{{\bf p}i}} \sin p_{ij} \sin k_{ij}$ for
$i,j=1,2,3$, where $V^\prime$ is the strength of the attractive
interaction. The pairing is then also $p$-wave type.

\vskip 0.15cm
{\sl Projected BdG Hamiltonian and Chern number.}-- We focus on
the weak pairing regime, where $|\mu|$ is much larger than
$|\Delta_\nu({\bf k})|$. In this regime one can safely ignore the
coupling between states separated with energy difference $\ge
|\mu|$. Consider $\mu>0$ (the results are similar for $\mu<0$ in the
checkerboard lattice), where we can project the BdG Hamiltonian 
into the subspace with only the $+$ band which crosses the Fermi
level. For simplicity, we consider the situation with $t_x=t_z$ in the
checkerboard lattice. The spectrum is then $\vep_{{\bf k}\pm} =
k^2(t_0\pm t_z)$ and $\vep_{{\bf k}\pm} = \frac{1}{8}|t|k^2(1\pm
1)$ in checkerboard and kagome lattices respectively. After the
projection, one obtains the Hamiltonian $H_{\rm
  PBdG}=\frac{1}{2}\sum_{\bf k} \Psi_P^\dagger({\bf 
  k})\mathcal{H}_{{\bf  k}}^{P} \Psi_P({\bf k})+\frac{1}{2}E_0$  
with $\Psi_P({\bf k})=(c_{{\bf k}+}, c_{-{\bf  k}+}^{\dagger})^T$ and
\be
{\cal H}_{\bf k}^{P} = \left[ \begin{array}{ccccccccc}
    \vep_{{\bf k}+} - \mu & & \Delta_{\rm eff}({\bf k}) \\
    \Delta_{\rm eff}^{\ast}({\bf k}) &  & -\vep_{{\bf k}+} + \mu  \\
\end{array} \right] ,
\label{pbdg}
\ee
where 
\be
\Delta_{\rm eff}({\bf k}) = \Delta_y({\bf k}) + \sum_{\pm}
\frac{1}{2}e^{\pm 2i\theta_{\bf k}} [\Delta_x({\bf k}) \mp  i\Delta_z({\bf k})] .
\ee
The spectrum is $\pm E_{\bf k}$ with $E_{\bf k}=\sqrt{\xi_{\bf k}^2 +
  |\Delta_{\rm eff}({\bf k})|^2}$ and $\xi_{\bf k}=\vep_{{\bf k}+} - \mu$. One finds again
that the singlet pairing does not contribute to $\Delta_{\rm eff}({\bf k})$ 
and the gap. The Chern number of the fully-gapped state is\cite{our} 
\be
N_C = \left. \frac{1}{2\pi}\int_0^{2\pi}d{\theta_{\bf k}} 
\partial_{\theta_{\bf k}}\theta_{\Delta}({\bf k})\right|_{\rm FS} 
\label{nc}
\ee
with $\theta_{\Delta}({\bf k})={\rm Arg}[\Delta_{\rm eff}({\bf k})]$,
which is just the winding number of the effective pairing $\Delta_{\rm
eff}({\bf k})$ at Fermi surface. A crucial observation 
is that the Chern number can only be {\em odd}, as only the triplet
pairing contribute to $\Delta_{\rm eff}({\bf k})$. Hence {\em all} the
fully-gapped pairing states are {\em non-Abelian} states\cite{our}.

\begin{figure}[htb]
\centerline{\includegraphics[height=3.1cm]{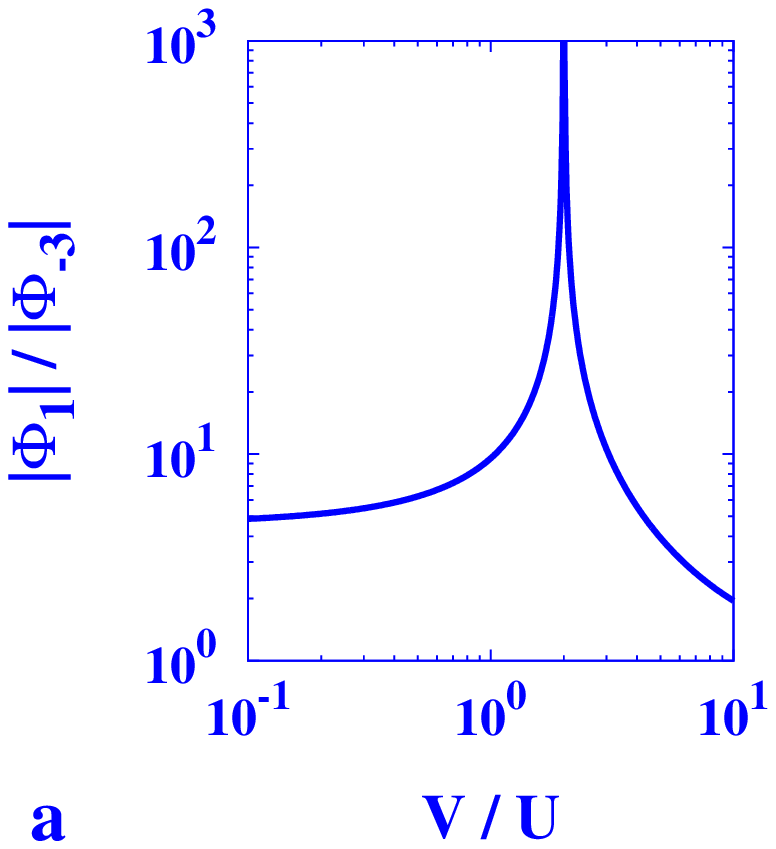}\includegraphics[height=3.1cm]{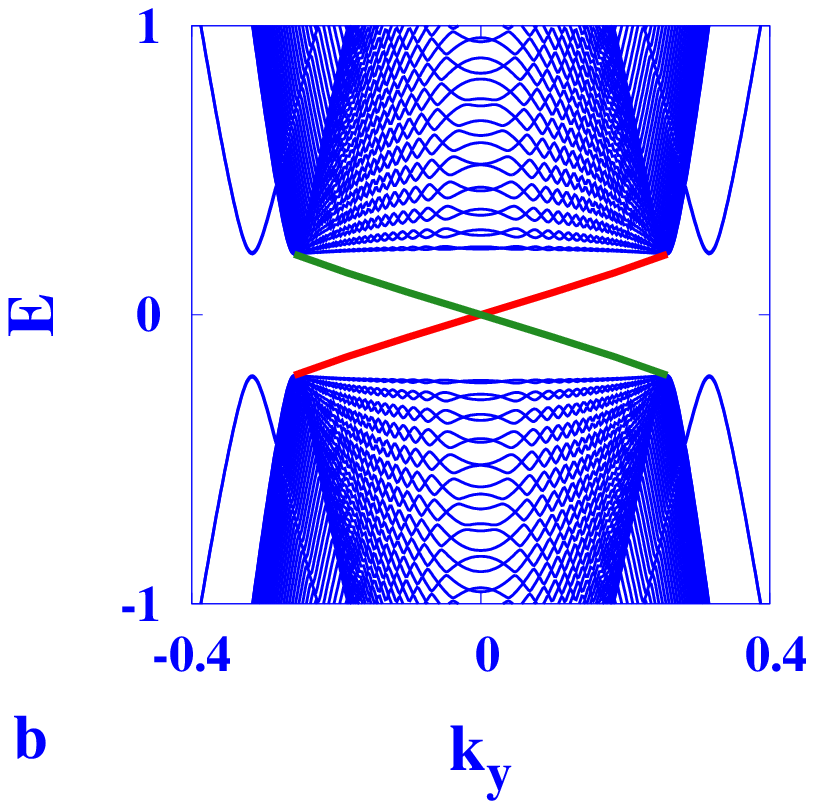}}
\caption{(Color online) In checkerboard lattice systems. (a) The ratio
  of the order parameters $|\Phi_1|/|\Phi_{-3}|$ as function of
  $V/U$. (b) Energy spectrum of the Bogoliubov quasi-particle
  versus $k_y$ in a stripe with periodic (open) boundary condition
  along the $y$ ($x$) -direction. The parameters are $t_0=10$,
  $t_z=t_x=20$, $\mu=2$, $V/2=U=1$, $g=0.95$. The width of
  the stripe along $x$ direction is $N_x=501$ unit cells.} 
\end{figure}

\vskip 0.15cm
{\sl Non-Abelian ground states.}-- To determine the ground states
we study the mean field free energy
\be
F = - k_BT\sum_{\bf k}\ln\left(2\cosh\frac{E_{\bf k}}{2k_BT}\right) 
+ \frac{1}{2}E_0 .
\ee
To facilitate the discussion, for checkerboard lattice we introduce
$g_{\nu\beta}\hspace{-0.1cm}=\hspace{-0.1cm}\sqrt{\frac{V_\nu}{2}}\sum_{\bf
  p}p_{-\beta}\ave{d_\nu^\dagger({\bf  p})}$ for $\beta=\pm$. This enables us to write $E_0 = g^2$ with
$g=\sqrt{\sum_{\nu,\beta} |g_{\nu\beta}|^2}$ being the ``pairing amplitude''
and 
$\Delta_{\rm eff}({\bf k}) = \frac{k\sqrt{U}}{2\sqrt{2}} \sum_{n}\Phi_ne^{in\theta_{\bf k}}$
with $n=\pm 1, \pm 3$, where $\Phi_{\pm 1}= 2 g_{y\pm} +  g_{x\mp} \mp i 
\sqrt{\frac{V}{2U}}g_{z\mp}$, $\Phi_{\pm 3}=g_{x\pm} \mp i
\sqrt{\frac{V}{2U}}g_{z\pm}$. By minimizing the free energy
numerically, we find that there are two possible ground states with
the same free energy: i) $\Phi_{1}$ and $\Phi_{-3}$ are finite with
$\Phi_{-1}=\Phi_3=0$, ii) $\Phi_{-1}$ and 
$\Phi_{3}$ are finite with $\Phi_{1}=\Phi_{-3}=0$. Both two states are
fully-gapped and break time-reversal symmetry. In fact, they are
time-reversal partners. Calculation indicates that the Chern
number of the two states are $N_C=\pm 1$ as $|\Phi_{1}|>|\Phi_{-3}|$ or
$|\Phi_{-1}|>|\Phi_{3}|$. In Fig.~2(a),
we plot the ratio $|\Phi_1|/|\Phi_{-3}|$ as a function of
$V/U$. It is seen that $|\Phi_{1}|>|\Phi_{-3}|$ in all the
parameter regime. At $V/U=2$, $\Phi_{-3}=0$ where
$\Delta_y:\Delta_x:\Delta_z=\frac{2}{\sqrt{6}}:\frac{1}{\sqrt{6}}:i\frac{1}{\sqrt{6}}$. Within
the mean field theory, for {\em all} $V/U$ the ground state is a
non-Abelian state with fully-gapped Fermi surface. We also plot the
spectrum of the Bogoliubov quasi-particles in Fig.~2(b) at $V/U=2$ as
an illustration of the edge states. From the figure it is seen that
there are two gapless chiral edge states which are localized at the
two boundaries separately. The self-consitent equation for the pairing
magnitude $g$ at $V/U=2$ is
$1=\frac{3}{8}\sum_{\bf k}\tanh(\frac{E_{\bf
    k}}{2k_BT})E_{\bf k}^{-1}k^2U$
with $E_{\bf k}=\sqrt{\xi_{\bf k}^2+\frac{3}{4}k^2Ug^2}$. From this we
obtain the transition temperature
$T_c \simeq
\frac{2\gamma}{\pi}\sqrt{\Lambda\mu}\exp[-\frac{16\pi
    (t_0+t_z)^2}{3U\mu} +\frac{\Lambda-\mu}{2\mu}] $,
where $\gamma$ is the Euler constant and $\Lambda$ is the high
energy cut-off. In 2D the SC/SF phase transition is
determined by the Kosterlitz-Thouless transition. However, in the weak
coupling regime, the Kosterlitz-Thouless transition temperature is
close to that obtained by the above approach\cite{cooper,miyake}.

For kagome lattice, we find that the ground state is also a
fully-gapped {\em non-Abelian} state where
$\Delta_{12}:\Delta_{13}:\Delta_{23}=1:e^{i\pi/3}:e^{i2\pi/3}$
(only $\Phi_1$ is finite and $N_C=1$) or
$\Delta_{12}:\Delta_{13}:\Delta_{23}=1:e^{-i\pi/3}:e^{-i2\pi/3}$ (only
$\Phi_{-1}$ is nonzero, $N_C=-1$). The transition temperature is
$T_c\simeq \frac{2\gamma}{\pi}\sqrt{\Lambda\mu} \exp[-\frac{3\pi
    |t|^2}{4V^\prime\mu} +\frac{\Lambda-\mu}{2\mu}]$.
In all those cases, the ground states are the fully-gapped states,
which is partly due to that such states gain more condensation energy
than the nodal states\cite{Anderson,Cheng}.

\vskip 0.15cm
{\sl Generalizations.}-- Here we generalize the property that {\em
  all} the singlet pairing instabilities are {\em absent} to more
classes of multi-orbital systems. As Eq.~(\ref{h0k}) is a general
description of two band touching, quite generally, the property
comes from the fact that ${\bf h}(-{\bf k})={\bf h}({\bf k})$. Hence
the property can also be realized in time-reversal symmetric systems
where the Fermi surface encloses a single two-band touching with any
{\em even} winding number. Such two-band touching is stable in systems
with time-reversal and space-inversion
symmetry\cite{Sun,our}. Moreover, the property holds even when such
band touching is gapped by finite $h_y({\bf k})$ via time-reversal
symmetry breaking, given that $h_y(-{\bf k})=h_y({\bf k})$ (space
inversion symmetric).
At finite $h_y({\bf k})$ the property remains to be true
even when $|h_0({\bf k})|>\sqrt{h_x^2({\bf k})+h_z^2({\bf k})}$ as
long as there is only one band crosses the Fermi level and another
is far away from it. It is also found\cite{our} that in those systems
{\em all} the fully-gapped states are {\em non-Abelian} states. In
analog with the checkerboard and kagome lattice systems, the
non-Abelian states are possibly the ground state in more general
systems as they have fully-gapped Fermi surface to gain more
condensation energy\cite{Anderson,Cheng}.

\vskip 0.15cm
{\sl Experimental realization and detections.}-- The checkerboard
optical lattice has been realized in experiments\cite{ch}. The scheme
to realize kagome optical lattice are proposed in Ref.~\cite{ka}.
The isotropic attractive interactions can be realized in Bose-Fermi
mixtures in deep lattices\cite{bfmix} or in polar-molecule gases
dressed by microwaves\cite{m,cooper}. Below we estimate the transition
temperature for polar-molecule gases in checkerboard lattices.
To have $T_c$ in the experimentally observable range, a moderately
strong interaction is needed\cite{cooper}. We take $t_0=0$,
$\Lambda=6t_z$ (the band width is $15t_z$), (e.g.) $\mu=0.8t_z$, and
$U=4t_z$, which gives $T_c\simeq 0.1t_z$ while the weak pairing condition
$|\Delta_{\rm eff}|_{T=0}\sim k_BT_c\ll |\mu|$ is still
satisfied. For a lattice with $t_z=200$~nK, $T_c\simeq 20$~nK. The
required interaction ($1600$~nK=$220$~KHz) is in the experimentally
accessible range for polar-molecules\cite{m}. The pairing symmetry and
the non-Abelian statistics in the proposed system can also be detected
experimentally via recently developed/proposed
techniques\cite{rmp,atom-qc,probing}.

{\sl Acknowledgements.}--I thank G. Conduit, S. Wu, J. Ruhman,
M. Gong, and Y. E. Krauss for fruitful discussions and
comments. This work was supported by the German Federal Ministry of
Education and Research (BMBF) within the framework of the
German-Israeli project cooperation (DIP) and by the Israel Science
Foundation (ISF).

\end{document}